\documentclass[twocolumn,showpacs,preprintnumbers,amsmath,amssymb]{revtex4}

\usepackage{graphicx}
\usepackage{graphicx,epstopdf}
\usepackage{dcolumn}
\usepackage{bm}

\begin{document}


\title{Restoration of oscillation in network of oscillators in presence of direct and indirect interactions }
\author{Soumen Majhi$^1$}
 \author{Bidesh K. Bera$^1$}
\author{Sourav K. Bhowmick$^2$}
\author{Dibakar Ghosh$^1$}
\email{diba.ghosh@gmail.com}
\affiliation{$^1$Physics and Applied Mathematics Unit, Indian Statistical Institute, Kolkata-700108, India\\
$^2$Department of Electronics, Asutosh College, Kolkata-700026, India}

\date{\today}

\begin{abstract} The suppression of oscillations in coupled systems may lead to several unwanted situations, which requires a suitable treatment to overcome the suppression. In this paper, we show that the environmental coupling in the presence of direct interaction, which can suppress oscillation even in a network of identical oscillators, can be modified by introducing a \emph{feedback factor} in the coupling scheme in order to restore the oscillation. We inspect how the introduction of the feedback factor helps to resurrect oscillation from various kind of death states. We numerically verify the resurrection of oscillations for two paradigmatic limit cycle systems, namely Landau-Stuart and Van der Pol oscillators and also in generic chaotic Lorenz oscillator. We also study the effect of parameter mismatch in the process of restoring oscillation for coupled oscillators.
\end{abstract}


\maketitle


\section{Introduction}
The interaction among oscillators in a system  can result in annihilation of oscillations and there are two dynamically and structurally different oscillation suppressing phenomena which are known as amplitude death (AD) \cite{ad} and oscillation death (OD) \cite{od}. In case of AD, oscillation is suppressed as all the coupled oscillators attain a homogeneous steady state (HSS) that was unstable in the uncoupled systems, whereas in the case of OD, oscillators populate coupling dependent stable inhomogeneous steady states (IHSS). The mechanisms leading to these two oscillation quenching phenomena are mainly detuning of oscillators under strong coupling \cite{m1, m2, m3}, conjugate coupling \cite{conj}, mean field coupling \cite{mf}, nonlinear coupling \cite{non}, additional repulsive link \cite{rep}, environmental coupling \cite{env1,env2} and also sufficient amount of delay in the coupling form \cite{Reddy&Sen.PRL.1998}. The diverse routes of transition from AD to OD have also been reported \cite{tr1,tr2,tr3,tr4,tr5,tr6}.

The revival or restoration of oscillation is a mechanism in which the rhythmicity of each oscillator in network of coupled oscillators is restored from a state of death without changing any intrinsic parameter of the individual oscillators. This has many possible applications due to the fact that stable oscillations are required for proper working of many physical, environmental and biological processes like brain waves \cite{new4}, cardiac and respiratory systems \cite{new3}, electric power generators \cite{new1}, El Ni\~{n}o/Southern Oscillation in Earth's ocean and atmosphere \cite{new2} and so on.  Usually oscillation quenched states are destructive and fatal in many real systems such as in all level of physiological process \cite{kou}, so revival of oscillations from death states using direct and environment interaction is very important. In the case of biological cells the environmental interaction is a surrounding medium \cite{katriel} where oscillatory reactions of the cell populations are taking place. On the other hand, the suppression and restoration of oscillation processes has an important role for epilepsy research where there is
	problem of electric stimulation of epileptic brain for discharge
	mitigation \cite{epinsc,epinscmthd,swd,nprneu,gs}. Applying delayed nonlinear feedback, the synchronized limit cycles, resembling clinical seizure EEG signals have also been suppressed \cite{epijtb}. Restoration of oscillation from death states of coupled systems is possible by introducing a processing delay \cite{rev1} in the coupling term. Chandrasekar \emph{et al.} \cite{rev2} used the feedback technique to restore the oscillations from the quenched state.\\
\par Very recently, general method for restoring oscillation from the death state has been proposed by Zou et al. \cite{rev3}, where they have introduced a \emph{feedback factor} and analyzed the efficiency of that approach using conjugate, dynamic and distributed delayed coupling and experimental evidence with coupled electrochemical reactions has also been given there. After that the usefulness of the aforesaid procedure has been justified using mean-field diffusive coupling \cite{rev4}. Revival of oscillations is also studied experimentally \cite{rev_chaos} in a nonlinear circuit using the processing delay in the interaction form,  also this scenario is experimentally verified using feedback factor in networks of electrochemical oscillators \cite{rev2_chaos} where oscillations are revived in anti-synchronization patterns.
But in absence of time-delay in the coupling term or parameter mismatch, simply diffusive coupling cannot induce AD. Again conjugate and dynamic coupling schemes have their own drawbacks, e.g., if we are dealing with a first order time-delay system then coupling with conjugate variables does not make any sense, demonstrating the fact that these schemes are not always quite general for inducing death and consequently the phenomenon of resurrection of oscillation from death state does not remain so general under the above mentioned schemes. On the other hand, in real world systems and many man made systems various type of complex phenomenon such as synchronization, amplitude death, oscillations death occur due to coupling through common external agency beside the presence of direct interactions. In this connection environmental coupling together with a direct interaction has been shown to be a  general mechanism for inducing suppression of oscillation in coupled systems taking almost all types of oscillators e.g., chaotic systems, periodic systems, model of neurons, time-delay systems etc. in \cite{env1} and also the effectiveness of this type of indirect interaction for quenching oscillation is possible in case of complex networks of oscillators \cite{env2}.

\par Inspired by the above facts, in this paper we consider a modified environmental scheme to make the phenomena more general, by introducing a \emph{feedback factor} and investigate its effect in networks of coupled oscillators. We show that by decreasing the value of feedback factor from unity, one can restore the rhythmicity from death state even for a network of identical and mismatched oscillators. Using detailed bifurcation analysis we explain that the death region in the parameter space of direct coupling and environmental coupling strength shrinks substantially if we decrease the value of feedback factor from unity. Previously, feedback controls are used for synchrony of coupled systems and also control diverse dynamics of the governed systems. There are various types of neuronal diseases where feedback control strategy is the most powerful treatment, for example in the case of Parkinson's diseases feedback control of deep brain simulation plays a crucial role to improve the efficacy and reduction of the side effect. For interference with electrical deep brain simulation spike-wave discharge provide a complementary tool and the real time detection of spike-wave discharge is experimentally verified in the ECoG of genetic rodent models \cite{swd}.
\par The remaining part of this paper is organized as follows. In Sec. II, we discuss the scenario of restoration of oscillations in coupled Landau-Stuart, Van der Pol and Lorenz oscillators. We also study the restoration of oscillations in a network of large number of identical oscillators. By detailed bifurcation analysis on coupled systems we characterize the death and revived oscillation states of the systems. Finally, we summarize our results in Sec. III.
 Scenario of resurrection of oscillation from various death states for coupled Landau-Stuart and Van der Pol oscillators interacting through only $y$ and both $x,y$ variables and the case of non-identical systems are given in Appendix A and B respectively.

\section{ RESTORATION OF OSCILLATION WITH MODIFIED SCHEME OF INTERACTION THROUGH AN ENVIRONMENT}
\subsection{Landau-Stuart oscillator}
In this section, we will demonstrate our scheme using paradigmatic model of Landau-Stuart (LS) oscillator which exhibits a stable limit cycle near supercritical Hopf bifurcation and has an unstable focus at origin. In the last two decades, Landau-Stuart oscillator has been used in the context of oscillation suppression and revival of oscillation using different configurations, as discussed earlier. So it will be interesting to explore our scheme using identical coupled Landau-Stuart oscillators. We consider $N$ environmentally coupled (all-to-all connected) Landau-Stuart oscillators in the form: \\
$$ \dot x_i=(1-{p_i}^2)x_i-\omega_i y_i+k\sum\limits_{j=1,j\neq{i}}^N (x_j-\alpha x_i)+\epsilon s ~~~~ \eqno{(1a)}$$
$$ \dot y_i=(1-{p_i}^2)y_i+\omega_i x_i,~~~i=1, 2, \cdot \cdot \cdot ,N,~~~~~~~~~~~~~~~ \eqno{(1b)}$$
$$ \dot s=- \gamma s-\epsilon \frac{(x_1+x_2+\cdot \cdot  \cdot  +x_N)}{N}, ~~~~~~~~~~~~~   \eqno{(1c)}$$
where $p_i^2=x_i^2+y_i^2$, $k$ is the strength of the direct diffusive coupling between the systems, $\epsilon$ is the strength of feedback coupling between the systems and the environment under a global interaction scenario, $\omega_i$'s are the intrinsic frequencies of individual LS oscillators. Here $\alpha$ $(0\le \alpha \le 1)$ is the feedback factor acting as a control parameter, where $\alpha=0$ represents the direct coupling in the presence of indirect interaction and $\alpha=1$ refers to the environmental coupling scheme  and $0<\alpha<1$ represents a bridge-linking between these two configurations.  The feedback factor $\alpha$ determines the proportion of the state of the $i$-th system to be received by all the other systems. $N$ is the number of oscillators in the network. Without loss of generality, we choose $\omega_i=\omega=2$ for $i=1,2, \cdot \cdot \cdot ,N$,  to make the oscillators identical. We consider the environment as a one-dimensional over-damped oscillator with damping parameter $\gamma.$ The $x$-components of LS oscillators are connected with each other and also with the environment. This coupling form breaks the rotational symmetry of the coupled oscillator, which is necessary for OD.
\par Let us begin with two identical diffusively coupled  Landau-Stuart oscillators together with an environmental interaction,
$$ \dot x_1=(1-{p_1}^2)x_1-\omega y_1+k(x_2-\alpha x_1)+\epsilon s ~~~~ \eqno{(2a)}$$
$$ \dot y_1=(1-{p_1}^2)y_1+\omega x_1 ~~~~~~~~~~~~~~~~~~~~~~~~~~~~~~ \eqno{(2b)}$$
$$ \dot x_2=(1-{p_2}^2)x_2-\omega y_2+k(x_1-\alpha x_2)+\epsilon s ~~~~ \eqno{(2c)}$$
$$ \dot y_2=(1-{p_2}^2)y_2+\omega x_2 ~~~~~~~~~~~~~~~~~~~~~~~~~~~~~~ \eqno{(2d)}$$
$$ \dot s=- \gamma s-\epsilon \frac{(x_1+x_2)}{2} ~~~~~~~~~~~~~~~~~~~~~~~~~~~~~~ \eqno{(2e)}$$
where $p_i^2=x_i^2+y_i^2, i=1, 2.$
The following are the equilibrium points of system (2): \\
(i) trivial homogeneous steady state (HSS), which is the origin (0,0,0,0,0), stabilization of which results in AD and additionally, two coupling-dependent nontrivial fixed points: \\
(ii) inhomogeneous steady state, $F_{IHSS}\equiv (a_1,b_1,-a_1,-b_1,0)$, where $a_1=\frac{-\omega b_1}{\omega^2+b_1^2 r_1}$ and $b_1=\pm\sqrt{\frac{(r_1-2 \omega^2)+\sqrt{r_1^2-4 \omega^2}}{2 r_1}}$ with $r_1=k(1+\alpha)$; stabilization of which results in OD and \\
(iii) non-trivial homogeneous steady state, $F_{NHSS}\equiv (a_2,b_2,a_2,b_2, -\frac{\epsilon a_2}{\gamma})$, where $a_2=\frac{-\gamma \omega b_2}{\gamma[\omega^2+k b_2^2 (\alpha-1)]+\epsilon^2 b_2^2}$ and $b_2=\pm\sqrt{\frac{P+\sqrt{Q}}{R}}$ with $P=\epsilon^4+2\gamma\epsilon^2 s_1+\gamma^2s_1^2-2\gamma^2\omega^2s_1-2\gamma\epsilon^2\omega^2,$
 $Q=\gamma^4s_1^4+6\gamma^2\epsilon^4s_1^2+\epsilon^8+4\gamma^3\epsilon^2s_1^3+4\gamma\epsilon^6s_1-4\gamma^4s_1^2\omega^2-8\gamma^3\omega^2\epsilon^2s_1
-4\epsilon^4\gamma^2\omega^2$ and $R=2(\gamma^2s_1^2+2\gamma\epsilon^2s_1+\epsilon^4)$ where $s_1=k(\alpha-1)$, stabilization of which results in non-trivial amplitude death (NAD) state whenever $\alpha=1$ (for which $F_{NHSS}$ would become independent of $k$  ). This newly observed NAD state not only has a nonzero homogeneous steady state, but more notably, in this state the system becomes bistable, which is shown later.

\begin{figure}[ht]
\centerline{
\includegraphics[scale=0.47]{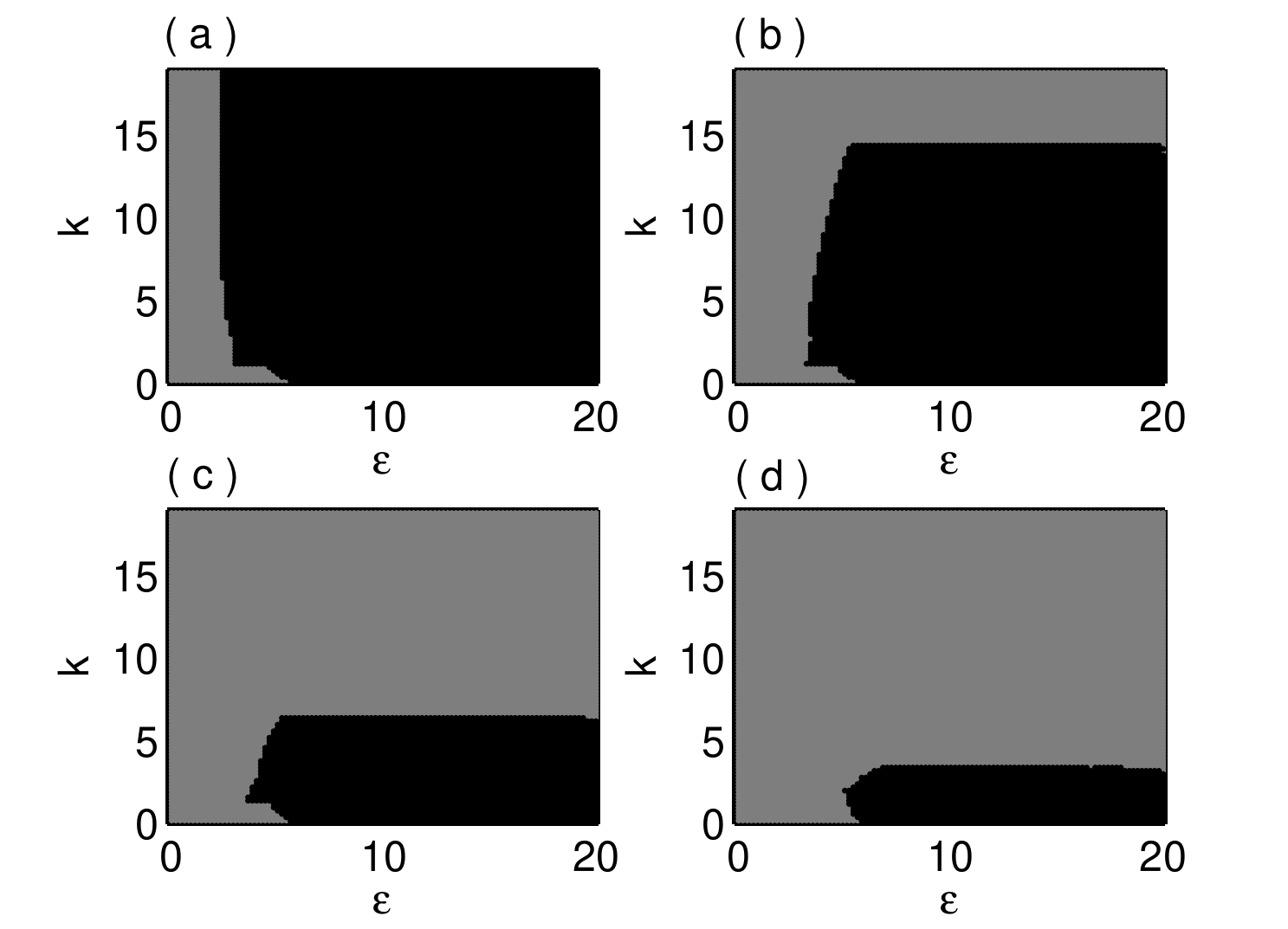}}
\caption{(Color online) Death regions (black color) and oscillation regions (gray color) in the $(\epsilon , k)$ parameter space for different values of feedback factor $\alpha$:  (a) $\alpha=1.0$, (b) $\alpha=0.8$, (c) $\alpha=0.6$,  and (d) $\alpha=0.1$ for fixed values of $\gamma=4.0, \omega=2.0$.}
\label{bds}
\end{figure}

\par For $\alpha=1$(i.e. the normal environmental scheme with direct interaction), the existence of death states and also transition scenario among them have been previously examined. In the present work we observe that even a feeble deviation of feedback parameter $\alpha$ from the normal environmental coupling drastically shrinks the death regions in the parameter space, which is depicted in Fig. 1, where the black and gray regions represent the death states (comprising of AD, OD and NAD states) and oscillation state respectively. For $\alpha=1$, large region of death is observed in $\epsilon-k$ phase plane in Fig. 1(a). If we decrease the value of feedback factor $\alpha$ to 0.8, then the death region shrinks (Fig. 1(b)) and oscillation states are revived where it was death for $\alpha=1.0$, as can be seen in Fig. 1(b).  Again by decreasing the value of $\alpha$  to $0.6$ and $0.1$, the death regions dramatically shrink and oscillation states are revived as in Figs. 1(c) and 1(d) respectively.
\par To confirm  death states, we calculate the average amplitude of the oscillators as the order parameter. The   average amplitude of the couple system (2) is defined as
$$d=\frac{d_1+d_2}{2},~~\mbox{where}~ d_i=\langle x_{i,max} \rangle -\langle x_{i,min}\rangle~~ \mbox{for}~ i=1,2 \eqno{(3)}$$
where $\langle\cdot \cdot \cdot \rangle$ represents the average over time. The parameter at which $d$ becomes zero is identified as the death state and non-zero for oscillatory state. The transition from oscillatory state to death state in the parameter plane $\alpha-\epsilon$ using the order parameter $d$ is shown in Fig. 2. We observe that at small values of feedback factor $\alpha$, there is no death region for any values of $\epsilon$ for $k=10.0.$

\begin{figure}[ht]
\centerline{
\includegraphics[scale=0.5]{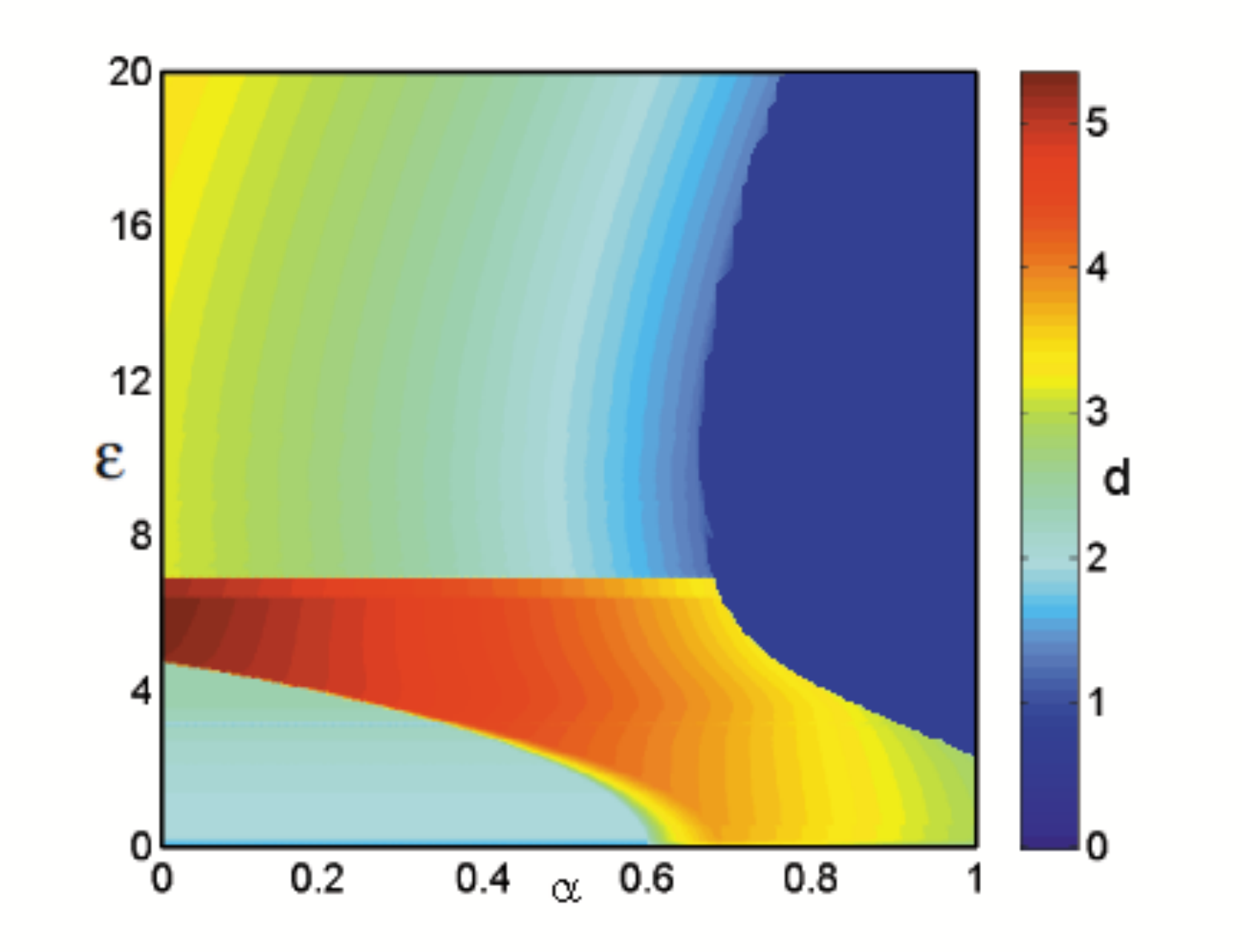}}
\caption{(Color online) Transition from oscillation state to death state in the parameter plane of environmental interaction strength $\epsilon$ and $\alpha$ in coupled Landau-Stuart oscillator (2) where average amplitude parameter $d$ is used as color bar. Death region is represented by deep blue (black) region and the rest of the region corresponds to the oscillatory state. Other parameters are same as in Fig. 1 with $k=10.0.$}
\label{bds}
\end{figure}

\par To reveal the complete scenario of the transition from intrinsic oscillatory dynamics to death and from death to restoration of oscillation, we draw the bifurcation diagram using XPPAUT \cite{xpp} and discuss in details. In Fig. 3(a) the transition from AD to OD state is observed for $\alpha=1.0$ and $\epsilon=3.2.$ In this case, the inhomogeneous steady state $F_{IHSS}$ does not depend on environmental interaction strength $\epsilon$ and solely depends on diffusive coupling strength $k$. So we plot the bifurcation diagram against the direct coupling strength $k$ keeping the environmental coupling strength $\epsilon$ as fixed. At lower values of diffusive coupling $k$, both the oscillators in the coupled system (2) are in oscillatory states and at $k=1.0$, AD occurs through supercritical Hopf bifurcation (HB) from stable limit cycle (SLC). This AD state is  stable for $1.0\le k\le 2.5$ and transition from AD to OD is observed for $k > 2.5$ through supercritical pitchfork bifurcation (PB) where the trivial fixed point i.e. AD state gets destabilized.  But interesting results are observed when we decrease the value of feedback factor to $\alpha=0.6$ in Fig. 3(b). Previously observed two types of quenched states (both AD and OD) are completely disappeared here and oscillations are revived via HB and unstable limit cycles (ULC) (blue line) transit to stable limit cycle through torus bifurcation. That means for a lower value of $\alpha$ than unity, the oscillation cessation state is  removed and stable limit cycle states are observed for all values of $ k\le 5.0$.
 \begin{figure}[ht]
\centerline{
\includegraphics[scale=0.48]{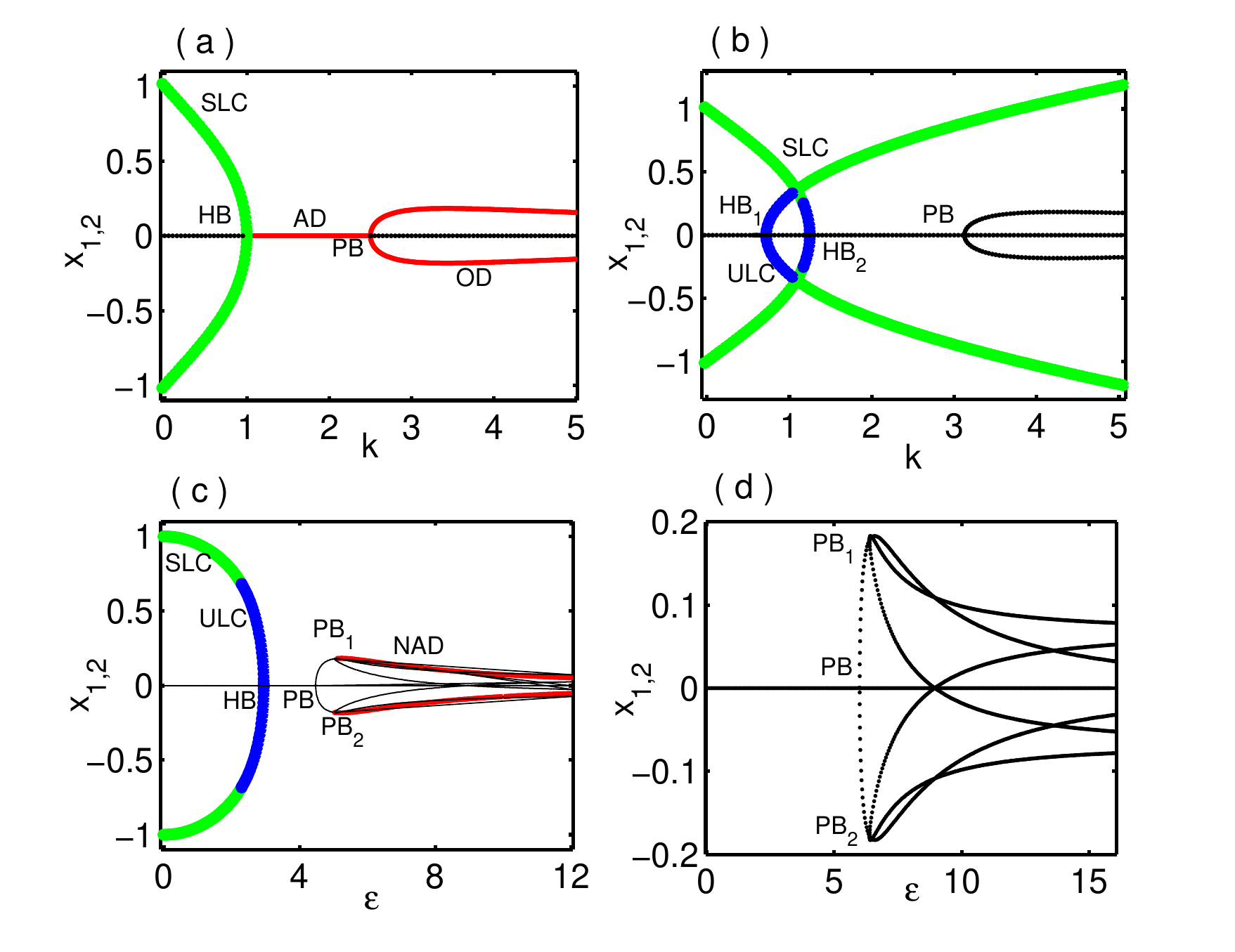}}
\caption{(Color online)  Two ($N=2$) environmentally coupled Landau-Stuart oscillators : Bifurcation diagrams with respect to $k$ for $(a) \alpha=1$ and $(b) \alpha=0.6$ at $\epsilon=3.2$, showing revival of oscillation from both AD and OD states. Bifurcation diagrams with respect to $\epsilon$ for (c) $\alpha=1$ and (d) $\alpha=0.6$ at $k=10.0$, showing revival of oscillation from NAD state. The red solid line is for the stable fixed point (AD, OD or NAD), black dotted lines are for unstable fixed points, the green open circle represents amplitude of stable limit cycle that emerges through Hopf bifurcation, and the blue open circle represents that of an unstable limit cycle.}
\label{bds}
\end{figure}

\par Next we explore the capability of the feedback factor $\alpha$ in retrieving oscillations in system (2) where NAD has been observed. The nonzero homogeneous steady state $F_{NHSS}$ depends on environmental coupling strength $\epsilon$, so we will study the behavior of the system by varying the environmental coupling strength $\epsilon$ for fixed value of direct coupling strength $k$. In Figs. 3(c) and 3(d), bifurcation diagrams are plotted with different values of $\epsilon$ for fixed value of $k=10.0$ for $\alpha=1.0$ and $\alpha=0.6$ respectively. In Fig. 3(c) we see that two state variables collapsed into a single non-zero fixed point (red lines) and populate to either the upper branch or the lower branch depending on the initial conditions ( due to bi-stability) after a certain threshold of $\epsilon$, namely $\epsilon=5.07$ via pitchfork bifurcation (PB1 and PB2) for $\alpha=1$. At $\alpha=0.6$, stability of NAD state is destroyed completely and oscillations are revived, as in Fig. 3(d). In fact in case of NAD, for any non-unit value of the limiting factor $ \alpha$ the nonzero homogeneous steady state $F_{NHSS}$ depends on both $\epsilon$ and $k$, as discussed earlier. So, as we decrease $\alpha$ from its unit value, the state of NAD immediately turns into a state of OD and a bit more decrement in $\alpha$ is leading to the restoration of oscillation of the coupled systems.

\par  We obtain similar results when two identical or mismatched LS oscillators are directly coupled through $y$ variables and both $x-y$ variables together with indirect interaction (The results are discussed in Appendix A) 

\begin{figure}[ht]
\centerline{
\includegraphics[scale=0.50]{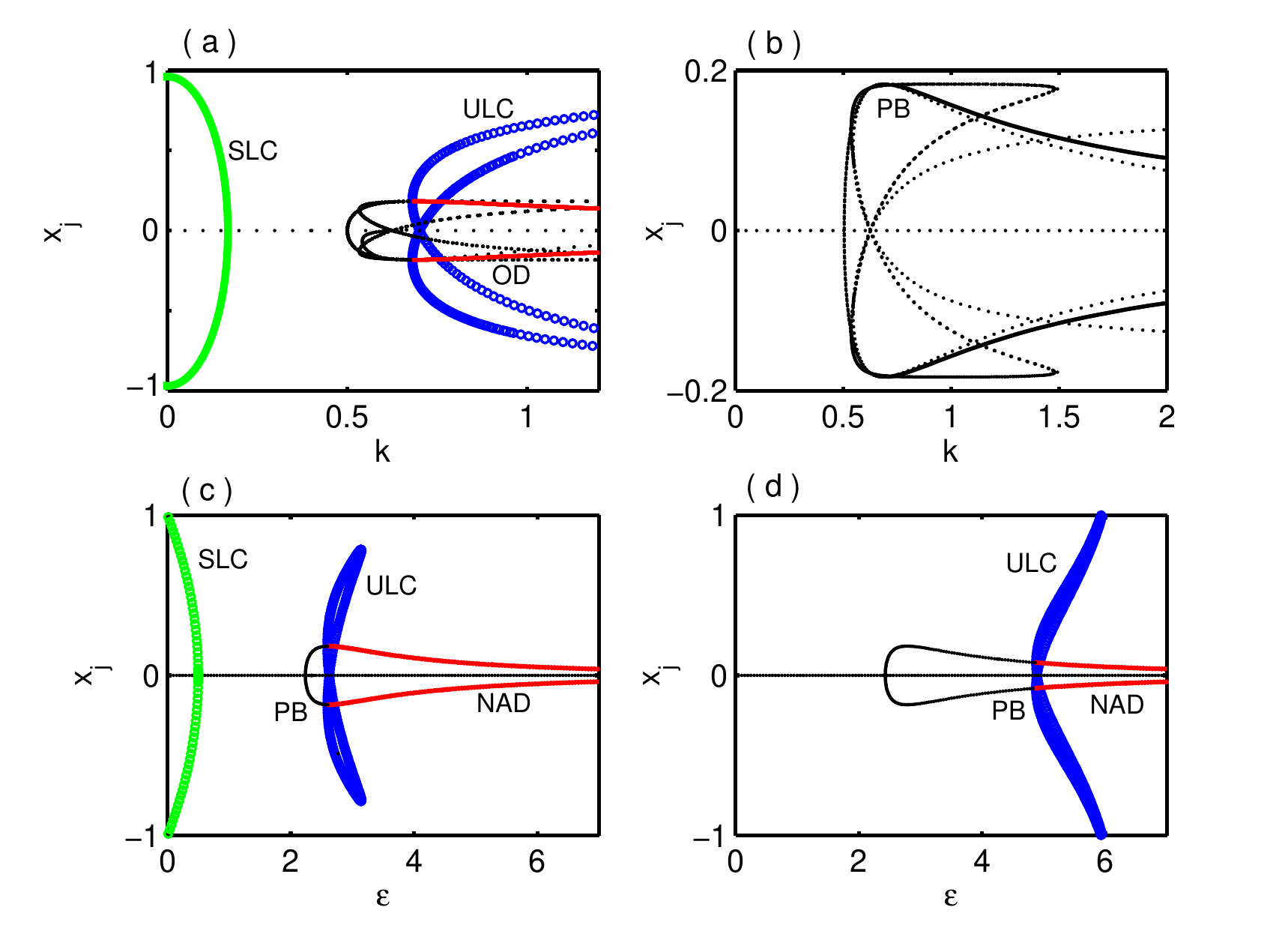}}
\caption{(Color online) Bifurcation diagram of a network ($N=10$) of coupled Landau-Stuart oscillators. The extrema values of the state variables $x_j (j=1,2,...,10)$  are plotted against the direct coupling strength $k$ while keeping the indirect coupling strength fixed at $\epsilon=2.5$ and feedback value at $\alpha=1$, that shows OD (a) and corresponding revival of oscillations for $\alpha =0.9$ (b). Similarly, (c) and (d) are respectively plotted with respect to $\epsilon$ for $\alpha=1$ and $\alpha=0.9$ with fixed values of $k=1.0$ that displays NAD and resurrected oscillation from NAD states.}
\label{10network}
\end{figure}

\begin{figure}[ht]
\centerline{\includegraphics[scale=0.460]{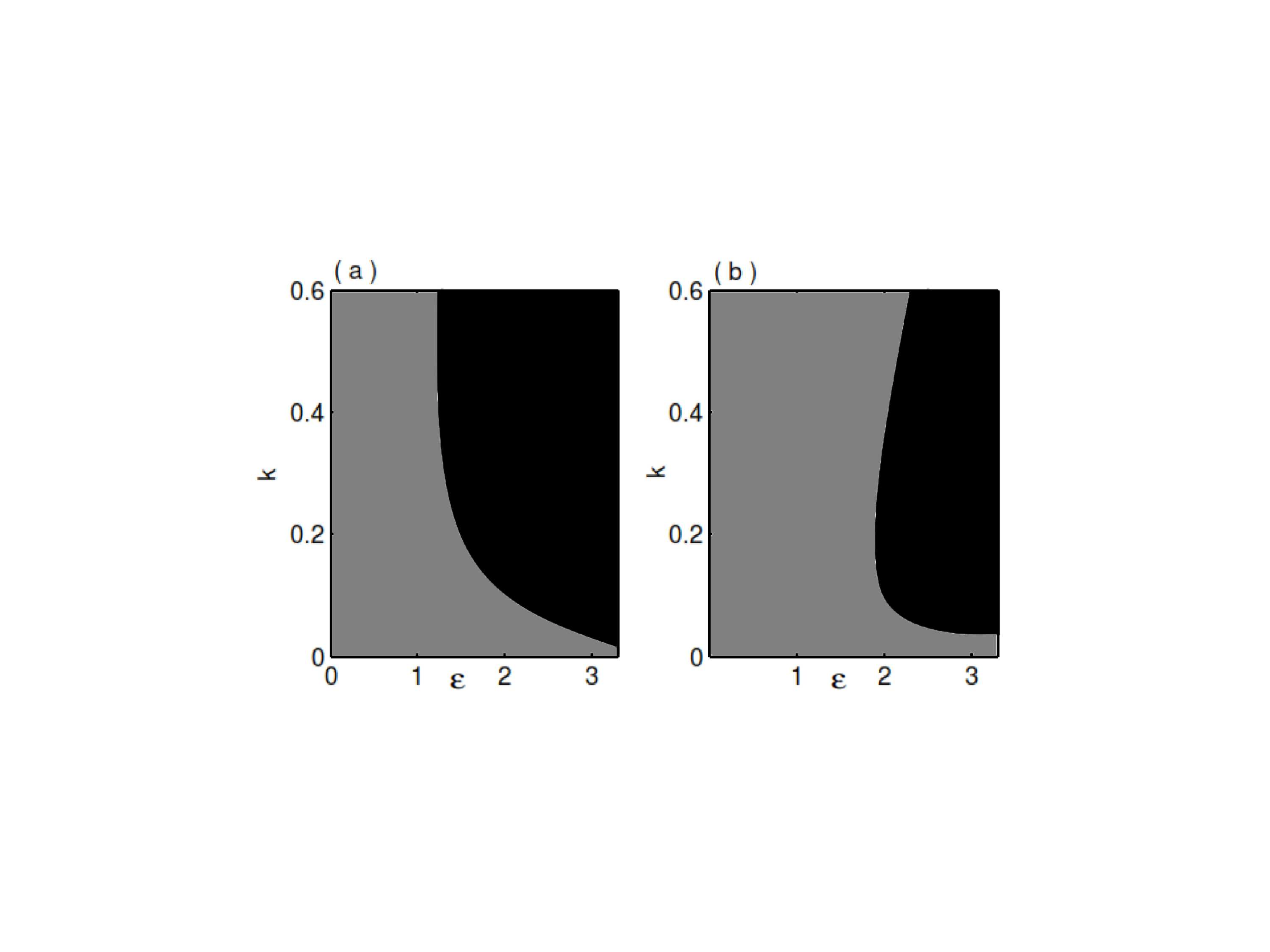}}
\caption{ Black and gray regions represent the corresponding death and oscillatory region in a network (N=100) of coupled Stuart-Landau oscillators. In (a) and (b) regions are plotted in $\epsilon - k$ plane for $\alpha=1$ and $\alpha=0.99$ respectively.}
\label{bds}
\end{figure}

\begin{figure}[ht]
\centerline{\includegraphics[scale=0.560]{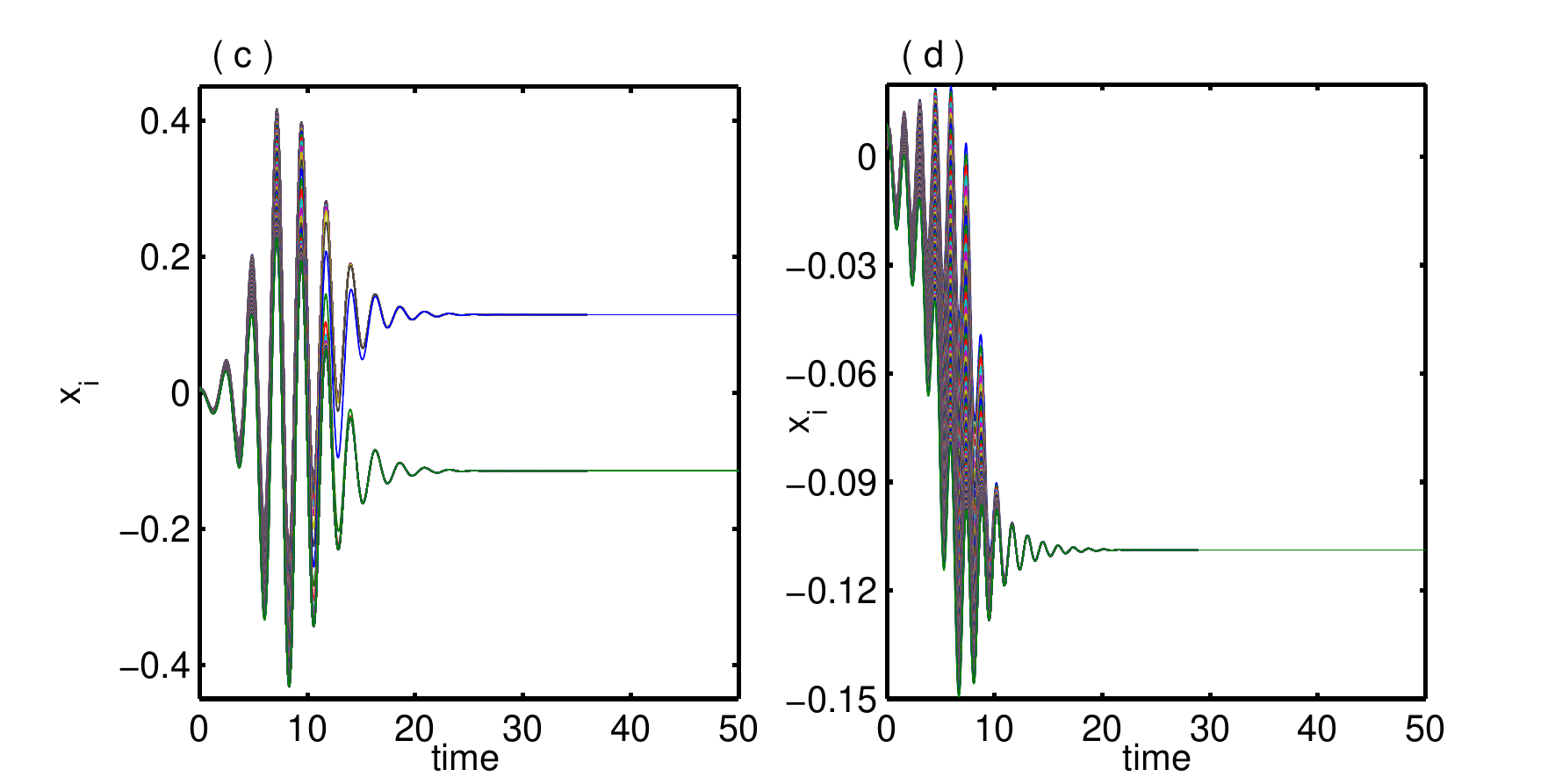}}
\caption{ The time series of the coupled network (N=100) are plotted for different steady states, (a) OD for $k=0.15, \epsilon =2.0$ and (b) NAD for $k=0.10, \epsilon =4.0$ with $\alpha=1.0$.} 
\label{bds}
\end{figure}

\begin{figure}[ht]
\centerline{
\includegraphics[scale=0.50]{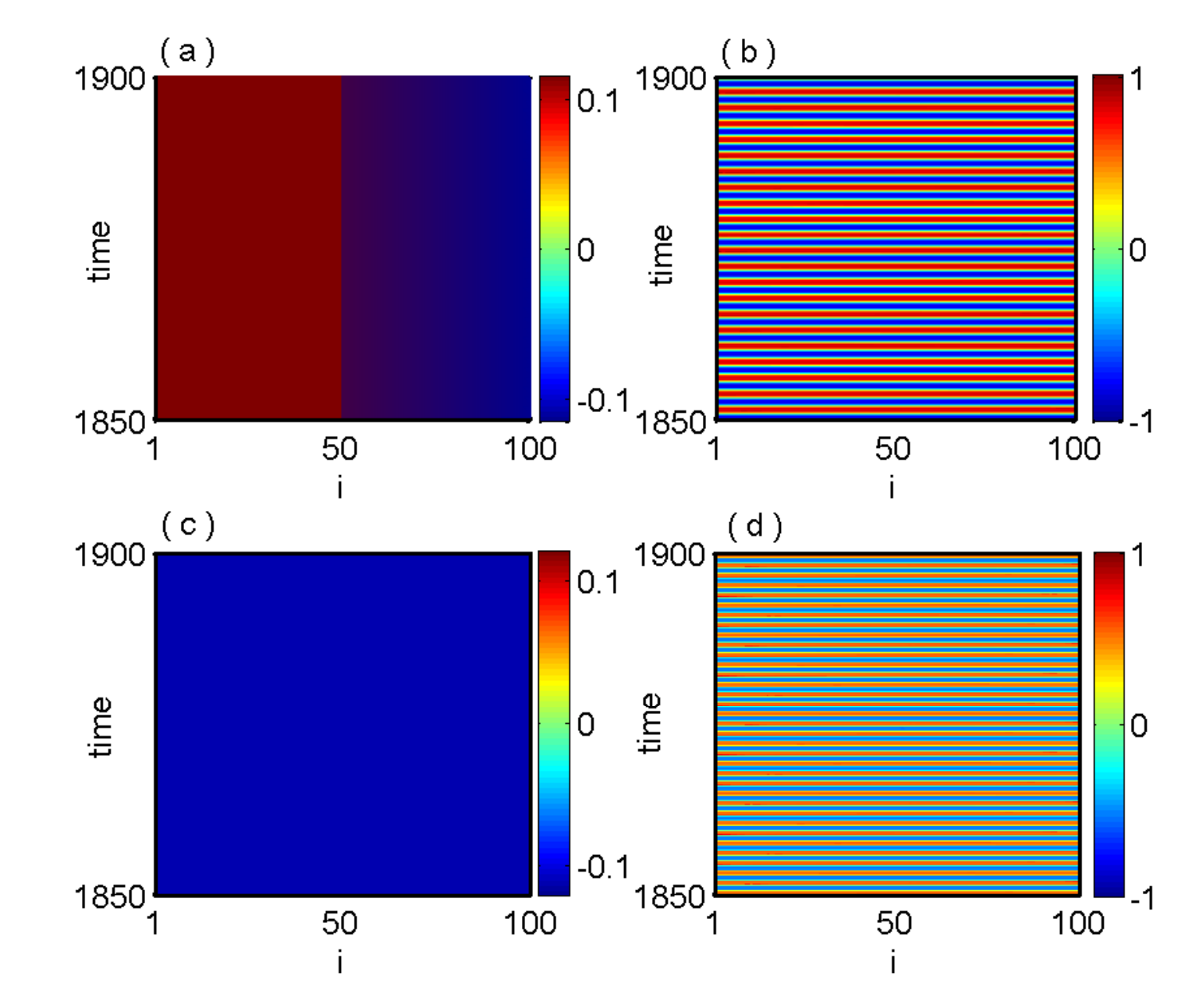}}
\caption{(Color on line) Network of $N=100$ coupled Landau-Stuart oscillators through common environment: spatio-temporal plots showing (a) OD  and (b) restoration of oscillation from OD for $\alpha=1.0$ and $\alpha=0.985$ respectively.  We fixed the other parameters as $k=0.15, \epsilon =2.0$. Space-time plots showing (c) NAD  and (d) resurrection of oscillation from NAD for $\alpha=1.0$ and $\alpha=0.930$ respectively. The other parameters are fixed at $\omega=2.0, \gamma=1.0, k=0.10, \epsilon =4.0.$  }
\label{bds}
\end{figure}

To see whether the study on revival of oscillation from death states for two coupled oscillators is valid also for large number of oscillators in a network, we consider a network of $N=100$ coupled Landau-Stuart oscillators (Eq. (1)) where all the nodes are globally connected with each other and also interacting through a common environment.  To check this point we first plot a bifurcation diagram (Fig. 4) for a small network (N=10) size with global coupling configuration. In Fig. 4(a) OD appears in a coupled network at a certain value of direct coupling strength $k=0.68$, for fixed indirect coupling strength $\epsilon=2.5$ and unit feedback value. Here among the $10$ oscillators, $5$ oscillators are populating to the upper branch while the next $5$ oscillators are approaching the lower branch of the IHSS. In Fig. 4(b) oscillations are revived for feeble deviation of feedback value from $\alpha=1.0$ to $\alpha=0.9$. The transition scenarios from oscillation to NAD and restoration of oscillations from that quenched state are shown in Figs. 4(c) and 4(d) respectively. In Fig. 4(c), NAD appears through PB at $\epsilon=2.6$ for fixed $k=1.0$ and $\alpha=1.0$ where all the oscillators either evolve to the upper branch or the lower branch of the NHSS, reflecting a bistable situation. Finally, in Fig. 4(d) oscillations are resurrected for $\alpha=0.9$. For further clarification we draw the phase space in Figs. 5(a) and 5(b) for a network of $N=100$ oscillators where black and gray regions represent the death and oscillatory states respectively for feedback values $\alpha=1.0$ and $\alpha=0.99$. Time series are shown in Figs. 6(a) and 6(b) corresponding to OD and NAD states for a network of $N=100$ coupled oscillators. In OD state, $50$ oscillators approach the upper state and the remaining $50$ oscillators move to the lower state due to bistability but in case of NAD, all the oscillators populate to the same branch. 
Fig. 7(a) shows the space-time plot of the network (1) of size $N=100$ depicting OD state (where the first $50$ oscillators populate to a branch of inhomogeneous steady state and the next $50$ oscillators populate to another branch of the steady state) for $k=0.15, \epsilon =2.0$. By decreasing the value of $\alpha$ (at $\alpha=0.985$), the OD state is revoked and synchronized rhythmicity is restored in the network (Fig. 7(b)).
Fig. 7(c) reflects NAD state (where all the $100$ oscillators populate to a single branch of the nontrivial homogeneous steady state) for $k=0.10, \epsilon =4.0$. For lower value of $\alpha$, the NAD state disappears (at $\alpha=0.930$) and synchronized oscillation is revived in the network (Fig. 7(d)). That means the decrement in the value of $\alpha$ results in restoration of synchronized oscillation from different types of death states even in a large network of coupled oscillators. This synchronized oscillations can be characterized by master stability function \cite{msf1,msf2,msf3}. 

 \subsection{Van der Pol oscillator}
Next, we explore the restoration process  from suppression states numerically in another limit cycle oscillators, namely Van der Pol oscillator.  Network of globally coupled Van der Pol oscillators in presence of common environmental interaction is described by the following equations as:\\
$$ \dot x_i=y_i+k\sum\limits_{j=1,j\neq{i}}^N (x_j-\alpha x_i)+\epsilon s~~~~ \eqno{(4a)}$$
$$ \dot y_i=a(1-x_i^2)y_i-x_i ~~~~ \eqno{(4b)}$$
$$ \dot s=- \gamma s-\epsilon \frac{(x_1+x_2+...+x_N)}{N} ~~~~ \eqno{(4c)}$$
where  $i=1,2,\cdot \cdot \cdot ,N$. Depending on the system parameter $a$, each uncoupled Van der Pol oscillator exhibits sinusoidal oscillation for $a<1$ and relaxation for $a>1$. Here we take $a=0.35$ to make sure that the oscillators are in limit cycle regime. $\gamma$ represents the damping constant of the environment.
 As stated earlier $k$ represents the direct diffusive coupling strength and $\epsilon $ is the   environmental interaction strength. We consider firstly the number of oscillators $N=2$ and find several fixed points of two coupled oscillators. The following are the fixed points: \\
(i) a trivial homogeneous steady state is the origin (0,0,0,0,0), stabilization of which results in (AD) and additionally, two coupling-dependent nontrivial fixed points\\
 (ii) $F_{IHSS}=(a_3,b_3,-a_3,-b_3,0)$, where $a_3=\frac{b_3}{r_2}$ and $b_3=\pm\sqrt{r_2^2-\frac{r_2}{a}}$, with $r_2=k(1+\alpha)$; stabilization of which results in OD and\\
(iii) $F_{NHSS}=(a_4,b_4,a_4,b_4,-\frac{\epsilon a_4}{\gamma})$, where $a_4=\frac{b_4}{s_2}$ and $b_4=\pm\sqrt{s_2^2-\frac{s_2}{a}}$, with $s_2=\frac{\epsilon^2}{\gamma}-k(1-\alpha)$; stabilization of which results in NAD, whenever $\alpha=1$ (for which $F_{NHSS}$ would become independent of $k$). We observe that the stabilization of newly created fixed point $F_{IHSS}$ in the OD state depends solely on diffusive coupling strength $k$ whatever be the value of feedback parameter $\alpha$. But in the NAD state fixed point $F_{NHSS}$ is independent of direct coupling strength $k$ only for $\alpha=1$. For slight deviation of $\alpha$ from unity, the fixed point $F_{NHSS}$ depends on both coupling strengths $k$ and $\epsilon$. After a certain value of $k$ and $\epsilon$ (which one is applicable), fixed points lose their stability and oscillations restore, which is discussed later.

\begin{figure}[ht]
\centerline{
\includegraphics[scale=0.55]{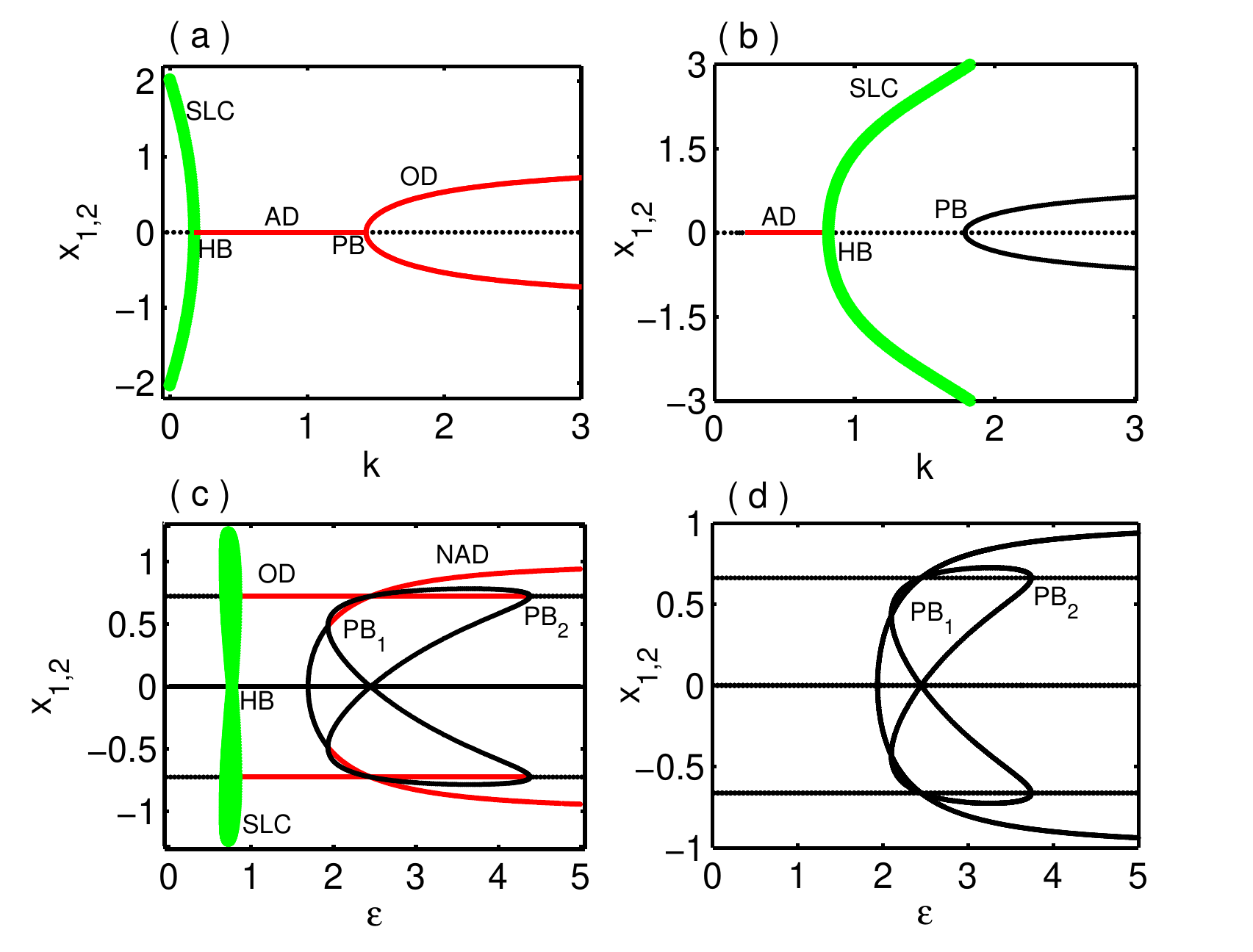}}
\caption{(Color online) Two ($N=2$) environmentally coupled Van der Pol oscillators: Bifurcation diagrams with respect to $k$ for (a) $\alpha=1$ and (b) $\alpha=0.6$ at $\epsilon=1.2$ respectively, showing revival of oscillation from both AD and OD states. Bifurcation diagrams with respect to $\epsilon$ for (c) $\alpha=1$ and (d) $\alpha=0.7$ at $k=3.0$ respectively, showing revival of oscillation from both NAD and OD states. The red solid line is for the stable fixed point, black dotted lines are for unstable fixed points and the green open circle represents amplitude of stable limit cycle that emerges through Hopf bifurcation.}
\label{bds}
\end{figure}

 \begin{figure}[ht]
   \centerline{
   \includegraphics[scale=0.50]{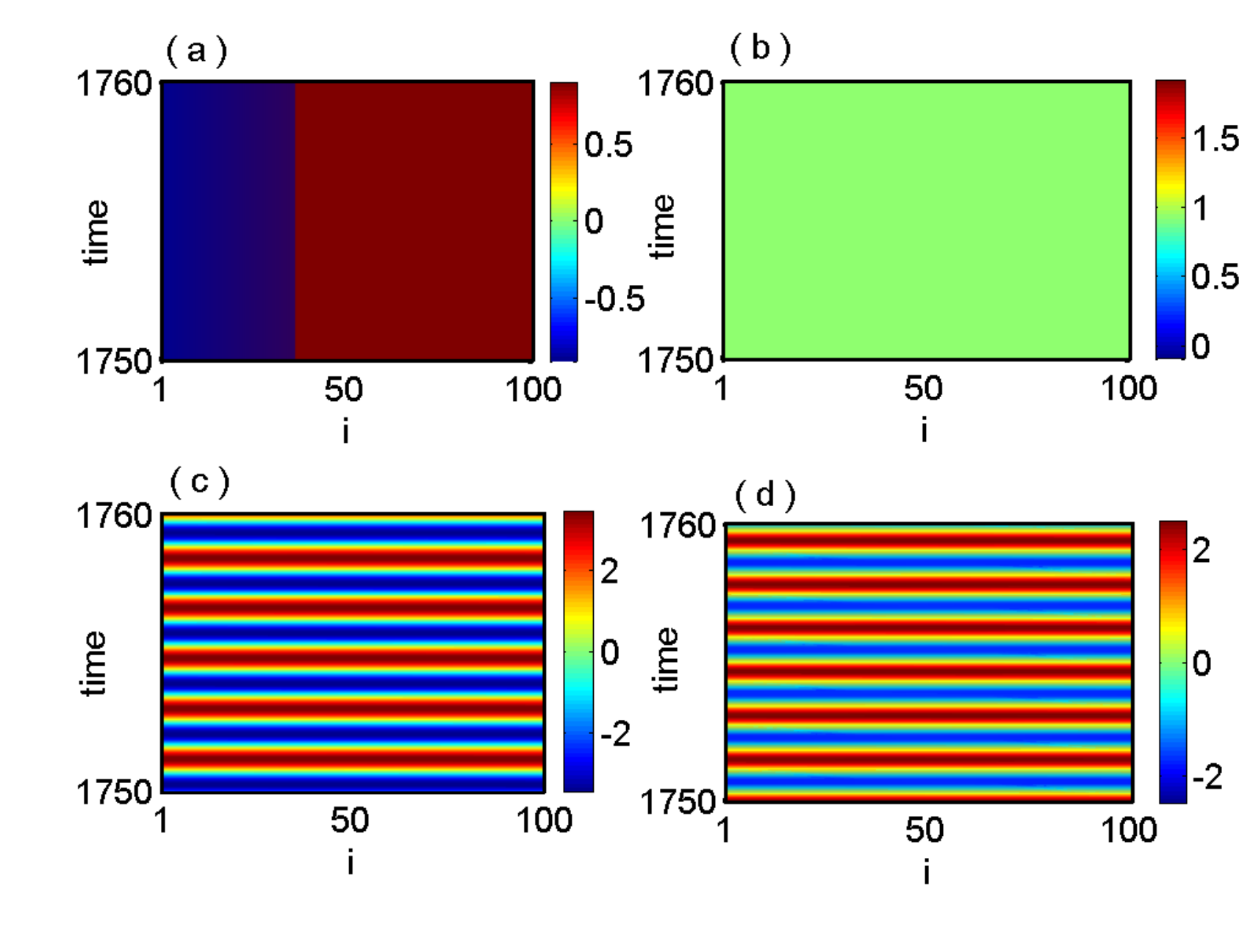}}
   \caption{(Color on line) The spatiotemporal plots for $N=100$ coupled Van der Pol oscillators: (a) OD state for $\epsilon=3.5, k=0.15$, (b) NAD state for $\epsilon=4.0, k=0.10$ where $\alpha=1.0.$ Revival of oscillations from (c) OD state for $\alpha=0.935$ and (d) NAD state for $\alpha=0.905$. Other parameters are $a=0.35, \gamma=1.0.$ }
   \label{bds}
   \end{figure}
 \par In Fig. 8(a), the bifurcation diagram shows the transition scenario from stable limit cycle to AD through supercritical Hopf bifurcation (HB) at $k=0.175$ and OD from AD via supercritical pitchfork bifurcation (PB) at $k=1.43$ for feedback parameter $\alpha=1.0$ and $\epsilon=1.2$ fixed. For lower value of $\alpha=0.6$, AD (red line) exists in smaller range of $k$ than that for $\alpha=1.0$ and OD completely disappears and oscillations are revived (Fig. 8(b)). Next we vary the environmental coupling strength $\epsilon$ for a fixed value of direct coupling strength $k=3.0$ and explore the effect of feedback factor $\alpha$ in retrieving oscillation from NAD. Figs. 8(c) and 8(d) show the NAD state (together with OD) for $\alpha=1$ and oscillation restoration state from this coexisting NAD and OD for $\alpha=0.7$ respectively. The stability of the fixed point $F_{NHSS}$ is completely destroyed in this regime by decreasing the value of $\alpha$ from unity. As discussed earlier in case of Landau-Stuart oscillator, here also in case of NAD, for any non-unit value of the limiting factor $ \alpha$ the nonzero homogeneous steady state $F_{NHSS}$ depends on both $\epsilon$ and $k$. So, as we decrease $\alpha$ from its unit value, the state of NAD rapidly turns into a state of OD and a further decrement in $\alpha$ is leading to the restoration of oscillation of the coupled systems.  We obtain similar results for coupling through $y$ and $xy$ variables with parameter mismatch and the results are shown in Appendix B.

\begin{figure}[ht]
   \centerline{
   \includegraphics[scale=0.50]{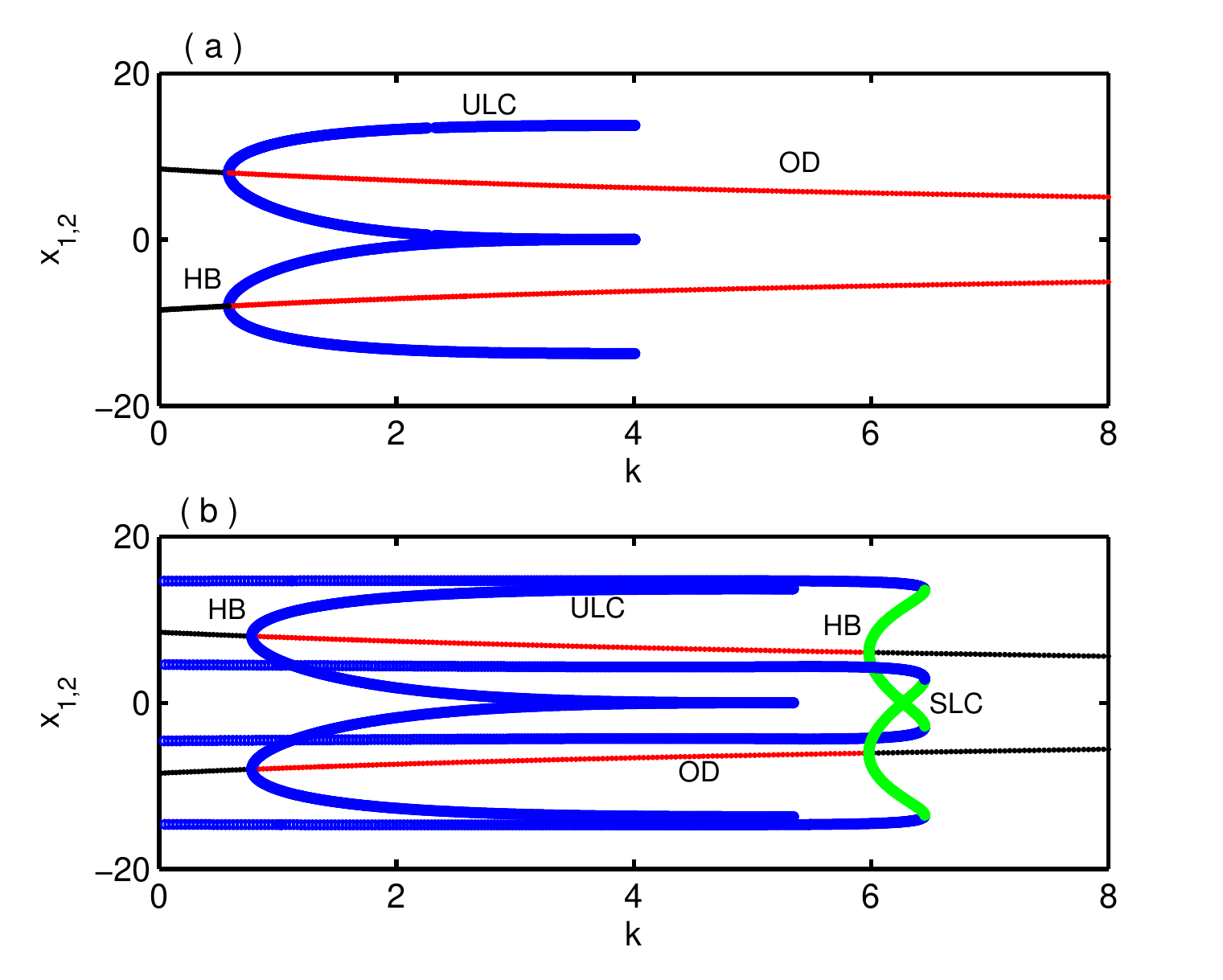}}
   \caption{(Color online) Two environmentally coupled Lorenz oscillators: Bifurcation diagrams with respect to $k$ for (a) $\alpha=1$ and (b) $\alpha=0.5$ at $\epsilon=10.0$ respectively, showing revival of oscillation from OD state. The red solid line is for the stable fixed point, black dotted lines are for unstable fixed points, the green open circle represents amplitude of stable limit cycle that emerges through Hopf bifurcation and the blue open circle represents that of an unstable limit cycle. Here $\gamma=1.0$.}
   \label{bds}
   \end{figure}

\par  Next we explored our result in a network of large number of Van der Pol oscillators taking $N=100$. For fixed values of diffusive coupling strength $k=0.15$ and environmental coupling strength $\epsilon=3.5$, oscillation death occurred for $\alpha=1.0$, shown in Fig. 9(a). In the case of OD state in the network, oscillators populate to two stable branches. Here $37$ oscillators populate to lower branch denoted by deep blue and the remaining $63$ oscillators converge to upper branch which is represented by deep red in Fig. 9(a). Again for $k=0.10$ and $\epsilon=4.0$, nontrivial amplitude death (NAD) occurred for $\alpha=1.0$ in Fig. 9(b) and in this case all the oscillators in the network converge to a single non-zero stable steady state. For a small deviation of the values of $\alpha$ from unity, oscillations are revived from both these OD and NAD states, shown in Figs. 9(c) and 9(d) for $\alpha=0.935$ and $\alpha=0.905$ respectively. In these states all the oscillators in the network show in-phase synchrony.

 \subsection{Lorenz oscillator}
 Finally,  to check the validity of restoration process from a suppression state in a system of two coupled chaotic oscillators, we extend our investigation in a chaotic Lorenz oscillator. Mathematical model of two coupled Lorenz oscillators in presence of common environmental interaction is described by the following equations as:\\
 $$ \dot x_{1,2}=\sigma (y_{1,2}-x_{1,2})+k(x_{2,1}-\alpha x_{1,2})+\epsilon s ~~~~ \eqno{(5a)}$$
 $$ \dot y_{1,2}=x_{1,2}(\rho-z_{1,2})-y_{1,2} ~~~~~~~~~~~~~~~~~~~~~~~~~~~~~~ \eqno{(5b)}$$
 $$ \dot z_{1,2}=x_{1,2} y_{1,2}-\beta z_{1,2} ~~~~~~~~~~~~~~~~~~~~~~~~~~~~~~ \eqno{(5c)}$$
 $$ \dot s=- \gamma s-\epsilon \frac{(x_1+x_2)}{2} ~~~~~~~~~~~~~~~~~~~~~~~~~~~~~~ \eqno{(5d)}$$
   The individual oscillators are in chaotic regime for the set of parameter values $\sigma=10, \rho=28, \beta=\frac{8}{3}$. 
   As stated earlier $k$ represents the direct diffusive coupling strength, $\epsilon $ is the   environmental interaction strength and $\alpha$ is a feedback parameter. In absence of interaction the individual oscillator has saddle at origin and another two unstable fixed points at $(\pm\sqrt{\beta(\rho-1)},\pm\sqrt{\beta(\rho-1)},\rho-1)$. The stabilization of these fixed points give rise to the amplitude death (AD). In the presence of direct as well as indirect interactions, the coupled systems converges to newly created fixed points. Followings are the coupling dependent fixed points: \\
   $F_{IHSS}=(a_5,b_5,c_5,-a_5,-b_5,c_5,0)$, where $a_5=\pm\sqrt{\frac {\sigma \beta c_5}{\sigma+r_3}}$, $b_5=\frac{\beta c_5}{a_5}$  and $c_5=\frac{\sigma(\rho-1)-r_3}{\sigma}$ with $r_3=k(1+\alpha)$; stabilization of which leads into the state of OD.\\
    \par In Fig. 10(a), the bifurcation diagram depicts the transition scenario from chaotic oscillation to OD through subcritical Hopf bifurcation (HB) at $k=0.59$ and continues to be in that state further for feedback parameter $\alpha=1.0$ and $\epsilon=10.0$ fixed. For lower value of $\alpha=0.5$, OD (red line) exists in smaller range of $k$ than that for $\alpha=1.0$ as shown in Fig. 10(b). Here chaotic oscillation gets restored after $k=5.98$ and the state of OD sustains only for $0.78\le k\le 5.98$. In fact, in this case also for $\alpha=0.5$ the two chaotic oscillators populate to two different symmetric branches of $F_{IHSS}$ (as given above) through subcritical Hopf bifurcation and remain in OD state upto $k=5.98$ and then periodic oscillation followed by chaotic oscillation is revived through supercritical Hopf bifurcation.    \\

\section{Conclusion}
We have studied the role of the feedback parameter $\alpha$ in reviving oscillation and maintaining rhythmicity in a network of oscillators in the presence of direct and indirect interactions by taking limit cycle as well as chaotic systems. The feedback parameter present in global coupling, which controls the rate of diffusion helps effectively in restoration of oscillation from different types of death states. The suppression of oscillations may lead to fatal system breakdown and an irrecoverable malfunctioning in many physiological, biological and physical systems. So, by controlling the feedback parameter $\alpha$ $(0\le \alpha \le 1)$ present in one of the most general mechanism for oscillation suppression, we have tried and have been able to resurrect oscillation state not only for two coupled systems but also for a network of large number of Landau-Start and Van-der Pol oscillators. We also study the effect of parameter mismatch and interaction among the oscillators through different pairs of variables while resurrecting oscillation from various types of death states. Lastly, we have analyzed the effectiveness of the mechanism for coupled Lorenz oscillators to resurrect oscillation from the state of OD.\\\\

{\bf{Acknowledgments:}}
Authors acknowledge the anonymous referees for their insightful suggestions. SKB acknowledges support by the DST, Govt. of India (Project no.YSS/2014/000687).\\\\\\

{\bf{Appendix A:} \\\\Restoration of oscillations scenarios from death states of coupled Landau-Stuart oscillators interacting through only $y$ and both $x,y$ variables:}
\\\\ The described coupling form for LS oscillators is given by
\\ $$ \dot x_{1,2}=(1-{p^2_{1,2}})x_{1,2}-\omega_{1,2} y_{1,2}+(1-\delta)k(x_{2,1}-\alpha x_{1,2})+(1-\delta)\epsilon s ~~~~ \eqno{(6a)}$$
$$ \dot y_{1,2}=(1-{p^2_{1,2}})y_{1,2}+\omega_{1,2} x_{1,2}+k(y_{2,1}-\alpha y_{1,2})+\epsilon s,~~~~~~~~~~~~~~~ \eqno{(6b)}$$
$$ \dot s=- \gamma s-\epsilon \frac{(1-\delta)(x_1+x_2)+(y_1+y_2)}{4-2\delta}, ~~~~~~~~~~~~~   \eqno{(6c)}$$
where $p_{1,2}^2=x_{1,2}^2+y_{1,2}^2 $ and all other parameters have the same meaning as in equation (2). Here $\delta=1$ and $\delta=0$ correspond to the cases of coupling through only $y$ variables and both $x-y$ variables respectively. Parameter mismatch in the form of $\omega_2=\omega_1+\Delta \omega $ where $\Delta \omega$ is the mismatch in the frequencies between the systems, is considered for the case of $\delta=1$.   
\begin{figure}[ht]
   \centerline{
   \includegraphics[scale=0.50]{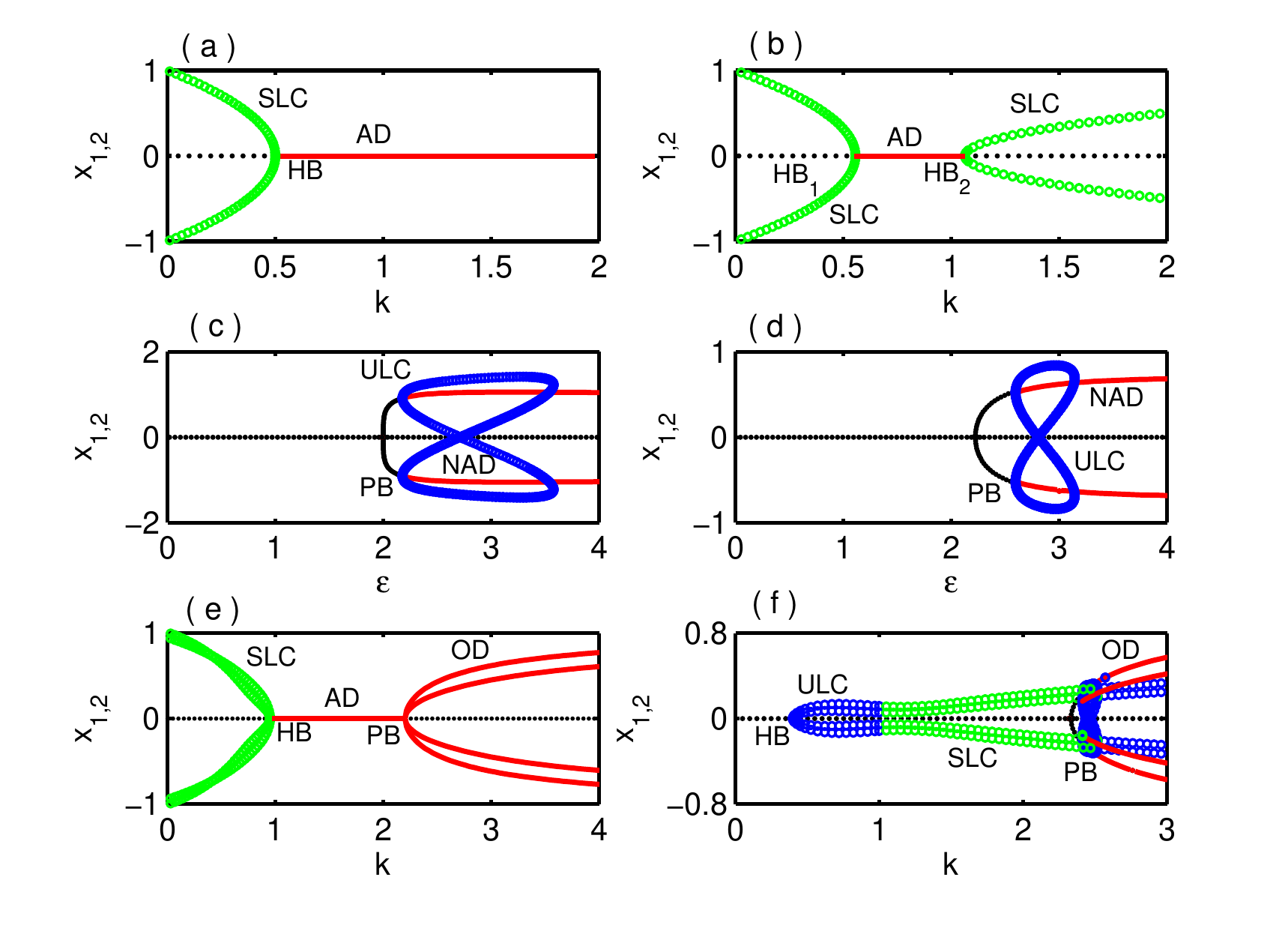}}
    \caption{(Color online)  Two ($N=2$) environmentally coupled Landau-Stuart oscillators : Bifurcation diagrams with respect to $k$ for (a) $\alpha=1$ and (b) $\alpha=0.8$ at $\epsilon=3.2$, showing revival of oscillation from AD state; with respect to $\epsilon$ for (c) $\alpha=1$ and (d) $\alpha=0.5$ at $k=2.0$, showing revival of oscillation from NAD state (for coupling through both $x$ and $y$ variables). Bifurcation diagrams with respect to $k$ for (e) $ \alpha=1$ and (f) $\alpha=0.9$ at $\epsilon=3.0$ with $\omega_1=1.7$ and $\Delta \omega=0.4$, showing revival of oscillation from  both AD and OD states (for coupling through only $y$ variables).}
   \label{fig9}
   \end{figure}
\\  Figure 11(a) shows the bifurcation diagram by varying direct coupling strength $k$ for $\epsilon=3.2, \alpha=1.0$ with $\delta=0$ i.e. two oscillators are coupled through both $x-y$ variables. As can be seen AD occurs there through a Hopf bifurcation at $k=0.50$ but if we decrease the value of $\alpha=0.8$, the AD for $1.06\le k \le 2.0$ disappears and revival of oscillation is observed in Fig. 11(b). Figure 11(c) shows the bifurcation diagram with respect to $\epsilon$ for $k=2.0$ and $\delta=0$ with $\alpha=1.0$, where NAD appears at $\epsilon=2.2$ through pitchfork bifurcation but for a decrement in the value of $\alpha$ to $\alpha=0.5$, NAD disappears for $2.2\le \epsilon \le 2.7$  and thus oscillation is restored in Fig. 11(d). Next we note that for coupled identical LS oscillators with coupling only through $y$ variables, the bifurcation scenarios and process of revival of oscillations are almost same as in the case of coupling through only $x$ variables as in Fig. 3. So here, we consider mismatch in the intrinsic frequencies of the oscillators (particularly, $\omega_1=1.7$ and $\Delta \omega=0.4$) while coupling is through only $y$ variables, i.e. $\delta=1$. With fixed $\epsilon=3.0$, we plot the bifurcation diagram by varying $k$ for $\alpha=1.0$ in Fig. 11(e) in which AD occurs at $k=1.0$ through Hopf bifurcation followed by appearance of OD through pitchfork bifurcation at $k=2.2$. A feeble deviation of $\alpha$ to $\alpha=0.9$ leads to resurrection of oscillation from both AD and OD, where AD completely disappears and OD remains for a smaller range of $k$ as in Fig. 11(f).\\  

 {\bf{Appendix B:} \\\\Resurrection of oscillations scenarios from quenched states of coupled Vander-pol oscillators interacting through $y-$ and $x,y$ both variables:}
\\\\ The dynamical equations of coupled VDP oscillators interacting through $y$ and $xy$ both the variables is given by
\\$$ \dot x_{1,2}=y_{1,2}+(1-\delta)k(x_{2,1}-\alpha x_{1,2})+(1-\delta)\epsilon s~~ \eqno{(7a)}$$
$$ \dot y_{1,2}=a_{1,2}(1-x_{1,2}^2)y_{1,2}-x_{1,2}+k(y_{2,1}-\alpha y_{1,2})+\epsilon s ~~ \eqno{(7b)}$$
$$ \dot s=- \gamma s-\epsilon \frac{(1-\delta)(x_1+x_2)+(y_1+y_2)}{4-2\delta} ~~ \eqno{(7c)}$$ 
All parameters bring the same meaning as in Eqn. no.(4). Here again, for coupling through only $y$ variables $\delta=1$ and coupling through both $x$, $y$ variables corresponds to $\delta=0$. Parameter mismatch in the form $a_2=a_1+\Delta a $ where $\Delta a$ is the amount of mismatch in between the systems, is opted for the case of $\delta=1$.
\begin{figure}[ht]
   \centerline{
   \includegraphics[scale=0.50]{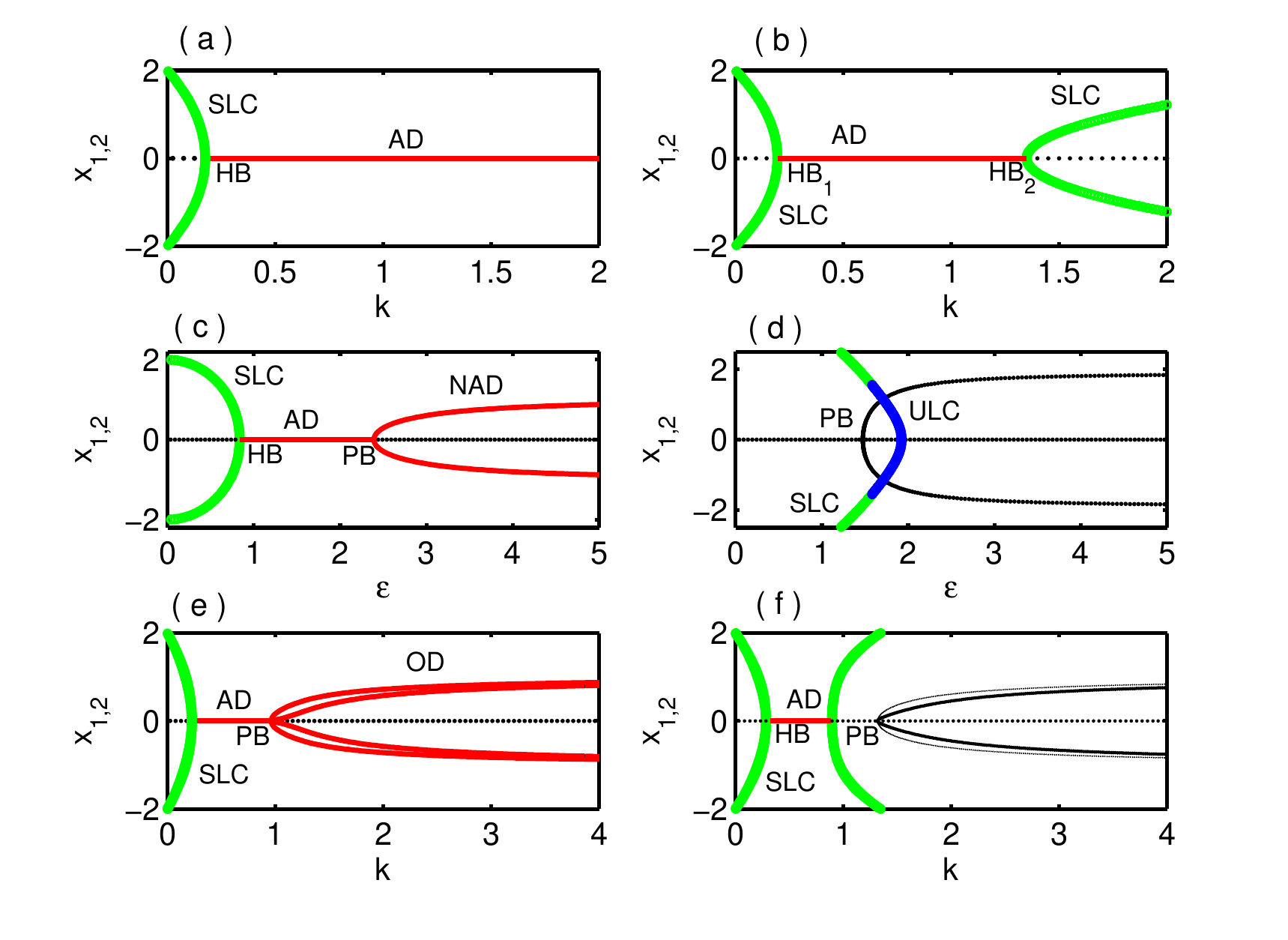}}
   \caption{(Color online)  Two ($N=2$) environmentally coupled Van-der Pol oscillators : Bifurcation diagrams with respect to $k$ for (a) $\alpha=1$ and (b) $\alpha=0.8$ at $\epsilon=1.5$, showing revival of oscillation from AD state (for coupling through only $y$ variables). Bifurcation diagrams with respect to $\epsilon$ for (c) $\alpha=1$ and (d) $\alpha=0.7$ at $k=1.5$, showing revival of oscillation from both AD and NAD states (for coupling through both $x$ and $y$ variables). Bifurcation diagrams with respect to $k$ for (e) $\alpha=1$ and (f) $\alpha=0.7$ at $\epsilon=1.3$ with $a_1=0.35$ and $a_2=0.55$, showing revival of oscillation from  both AD and OD states (for coupling through only $x$ variables).}
   \label{fig10}
   \end{figure}
   
In Fig. 12(a), we plot the bifurcation diagram by varying direct coupling strength $k$ for $\epsilon=1.5, \alpha=1.0$ with $\delta=1$ i.e. two oscillators are coupled through only $y$ variables. We observe that AD occurs here at $k=0.22$ through a Hopf bifurcation but as we decrease the value of $\alpha=0.8$, the AD for $1.35\le k \le 2.0$ disappears and oscillation is revived again through Hopf bifurcation in Fig. 12(b). Figure 12(c) shows the bifurcation diagram with respect to $\epsilon$ for $k=1.5$ and $\delta=0$ with $\alpha=1.0$, where AD and NAD appear at $\epsilon=0.835$ and $\epsilon=2.4$ through Hopf and pitchfork bifurcations respectively but decreasing the value of $\alpha$ to $\alpha=0.7$, both AD and NAD disappear and oscillation is restored for all values of $\epsilon$ in Fig. 12(d). Finally, we consider mismatch in the intrinsic parameter $a$ of the oscillators (particularly, $a_1=0.35$ and $\Delta a=0.2$) while coupling is through only $x$ variables. Fixing $\epsilon=1.3$, we plot the bifurcation diagram by varying $k$ for $\alpha=1.0$ in Fig. 12(e) where AD occurs at $k=0.23$ through Hopf bifurcation followed by appearance of OD through pitchfork bifurcation at $k=0.96$. Deviation of $\alpha$ to $\alpha=0.7$ leads to resurrection of oscillation from both AD and OD, where OD completely disappears and AD remains for a smaller range of $k$ as in Fig. 12(f).\\\\

\end{document}